\documentclass[11pt]{article}
\usepackage{amsfonts}
\usepackage{amsmath}
 \usepackage{amssymb}
 \newfont{\bbbold}{msbm10}

 \def\bbC{\mbox{\bbbold C}}

 \def\bbF{\mbox{\bbbold F}}

 \def\bbP{\mbox{\bbbold P}}

 \def\bbZ{\mbox{\bbbold Z}}

 \newfont{\goth}{eufm10 scaled \magstep1}

 \def\a{\alpha}
 \def\b{\beta}
 \def\c{\gamma}
 \def\d{\delta}
 \def\e{\epsilon}
 \def\f{\phi}\def\F{\Phi}
 
 \def\i{\iota}

 \def\m{\mu}

 \def\t{\tau}
 \def\th{\theta}
 
 \def\O{\Omega}
 
 \def\adt{\dot \alpha}
 \def\bdt{\dot \beta}
 \def\cdt{\dot\gamma}
 \def\ddt{\dot\delta}
 \def\be{\begin{equation}}\def\ee{\end{equation}}
 \def\bea{\begin{eqnarray}}\def\eea{\end{eqnarray}}
 \def\ba{\begin{array}}\def\ea{\end{array}}
 \def\del{\partial}
 \def\half{{1\over2}}\def\qu{{1\over4}}

 {}
 \def\nn{\nonumber}
 \def\bea{\begin{eqnarray}}
 \def\ba{\begin{array}}\def\ea{\end{array}}
 \def\eea{\end{eqnarray}}

 \def\tr{{\rm tr}}

 \def\xz{\times}

\begin{document}

\thispagestyle{empty}

 \hfill{KCL-MTH-04-12}

  \hfill{\today}

 \vspace{30pt}

 \begin{center}
 {\Large{\bf $R^4$ terms in supergravity and M-theory}}
 \vspace{40pt}

 {P.S. Howe} \vskip .5cm {Department of Mathematics}
 \vskip .5cm {King's College, London} \vspace{15pt}

 \vspace{60pt}

 \end{center}

 {\bf Abstract}

Higher-order invariants and their r\^ole as possible
counterterms for supergravity theories are reviewed. It is argued that $N=8$ supergravity will diverge at 5 loops. The 
construction of $R^4$ superinvariants 
in string
theory and M-theory is discussed.

 {\vfill\leftline{}\vfill \vskip  10pt

 \baselineskip=15pt \pagebreak \setcounter{page}{1}

\section{Introduction}

The title of this contribution to the Deserfest is an appropriate
 one in view of the fact that the first paper on three-loop
 ($R^4$)
 counterterms in supergravity was written by Stanley Deser, together
 with John Kay and Kelly Stelle, in 1979 \cite{Deser:1977nt}. It will be recalled that
 supergravity had been shown to be on-shell finite at one and two
 loops \cite{Grisaru:1976nn} and there were hopes that this would persist at higher loop
 orders, but these were somewhat spoiled by the above paper, on
 $N=1$ supergravity, and by a follow-up which reported a similar
 result for $N=2$ \cite{Deser:1978br}.  Any residual hopes that the maximally
 supersymmetric $N=8$ supergravity theory in four dimensions would
 have special properties were removed by the observation that
 counterterms can easily be constructed as full superspace
 integrals, at seven loops in the linearised theory, and at eight
 loops if one wishes to preserve all the symmetries of the full
 non-linear theory including $E_7$ \cite{Howe:1980th}. Subsequently, a linearised
 three-loop $N=8$ invariant was also found \cite{Kallosh:1980fi}. In view of the
 difficulty of carrying out high loop calculations in quantum gravity, and of
 the success of string theory and M-theory, the subject has receded
 into the background over the years but interest in it has
 recently been reawakened by the development of new calculational
 techniques \cite{Bern:1998ug}. The implications of this work and its relation to the
 predictions of superspace power-counting arguments,
 both in conventional superspace and in harmonic superspace, will be the
 subject of the next section of the paper.

 Notwithstanding these recent developments, $R^4$ and other
 higher order supersymmetric invariants are nowadays of most
 interest in the context of effective field theory actions in string
 theory and M-theory. Such terms could be interesting both
 theoretically, especially in the case of M-theory, and in
 applications to solutions and to beyond leading-order tests of
 various implications of duality. For such applications one would
 ideally like to know the complete bosonic part of the effective
 action. However, these terms seem to be extremely difficult to
 calculate systematically as will be seen below. Some partial results
 obtained from string theory and supersymmetry will be
 outlined in section 3 and the final section of the paper will be
 devoted to a brief exposition of a superspace approach to the
 problem in eleven dimensions \cite{Howe:2003cy}. This work seems to indicate that there is
 a unique invariant at this order which is supersymmetric and
 which is compatible with the M-theoretic input of five-brane anomaly
 cancellation.

\section{Supergravity counterterms}

The $D=4,N=8$ three-loop counterterm was first constructed in \cite{Kallosh:1980fi}.
A manifestly supersymmetric and $SU(8)$ invariant version was
given in \cite{Howe:1981xy} as an example of a superaction - an integral over
superspace which involves integrating over fewer than the total
number of odd coordinates. This type of integration, using
conventional superspace and measures which carry $SU(N)$
representations, turns out to be equivalent to integration in
certain harmonic superspaces  where the number of odd
coordinates is reduced \cite{Hartwell:1994rp}. The three-loop counterterm has a very
simple form in this language.

In $D=4$ $N$-extended supersymmetry harmonic superspaces are
obtained from Minkowski superspace by adjoining to the latter a
coset space of the internal $SU(N)$ symmetry group, chosen to be
complex \cite{Galperin:1984av}. This coset can be thought of as parametrising sets of
mutually anticommuting covariant derivatives ($Ds$ and $\bar Ds$).
The simplest example is in $N=2$ where we can select one $D_\a$
and one $\bar D_{\adt}$ to anticommute\footnote{$\a$ and $\adt$
denote two-component spinor indices}, and the ways this can be
done are parametrised by the two-sphere  $\bbC \bbP^1$. For
higher values of $N$ there are more ways of choosing such sets of
derivatives and therefore many different types of Grassmann
analyticity (or generalised chirality) constraints that one can
impose on superfields. In order for covariance with respect to the
R-symmetry group to be maintained such G-analytic superfields must
be allowed to depend on the coordinates of the coset space. It
turns out that the G-analyticity constraints are compatible with
ordinary (harmonic) analyticity on the coset space and that many
field strength superfields can be described by superfields which
are analytic in both senses. Harmonic analyticity ensures that
such fields have short harmonic expansions.

To be more explicit, let $i=1\ldots N$ and $I=1 \ldots N$ denote
internal indices which are to be acted on by $SU(N)$ and the
isotropy group respectively. We split $I$ into three,
$I=(r,R,r')$, where the ranges $1\ldots p$ and $N-(q+1)\ldots N$
of $r$ and $r'$ cannot be greater than $N/2$; we write $u\in
SU(N)$ as $u_I{}^i=(u_r{}^i,u_R{}^i,u_{r'}{}^i)$ and similarly for
the inverse element $(u^{-1})_i{}^I$. This splitting is clearly
preserved by the isotropy group $S(U(p)\xz U(N-(p+q))\xz U(q))$,
the coset space defined by this group being the flag space
$\bbF_{p,N-q}$ of $p$-planes within $(N-q)$-planes in $\bbC^N$.
Let

 \be D_{\a I}=u_I{}^i D_{\a i};\qquad \bar D_{\adt}^I =
 (u^{-1})_i{}^I\bar D_{\adt}^i\ ,
 \ee
then the derivatives $D_{\a r}$ and $\bar D_{\adt}^{r'}$ are
mutually anti-commuting. Superfields which are annihilated by
these derivatives are said to be G-analytic of type $(p,q)$; the
associated harmonic superspace is known as $(N,p,q)$ harmonic
superspace.

This formalism can be applied to $N=8$ supergravity. The
supergravity multiplet is described by the linearised field
strength $W_{ijkl},\ i=1\ldots 8$. The superfield $W_{ijkl}$ is
totally antisymmetric and transforms under the seventy-dimensional
real representation of $SU(8)$. It obeys the constraints

 \bea
 \bar W^{ijkl}&=&\frac{1}{4!}\e^{ijklmnpq}W_{mnpq}\\
 D_{\a i}W_{jklm}&=&D_{\a [i}W_{jklm]}\\
 \bar D_{\adt}^i W_{jklm}&=&-\frac{4}{5}\d^i_{[j}\bar D^n_{\adt} W_{klm]n}
 \eea
the third of which follows from the other two. Note that this
superfield defines an ultra-short superconformal multiplet.

The same multiplet can be described by an analytic superfield $W$
on $(8,4,4)$ harmonic superspace:

 \be
 W:=\e^{rstu} u_r{}^i u_s{}^j u_t{}^k u_s{}^l W_{ijkl}
 \ee
It is not difficult to see that the constraints

 \be
 D_{\a r}W=\bar D_{\adt}^{r'}W=0,\qquad r=1\ldots 4,\
 r'=5,\ldots 8\, ,
 \ee
together with analyticity on the internal coset, are equivalent to
the differential constraints above, while the reality condition can
also be formulated in harmonic superspace.

Since $W$ depends on only half of the odd coordinates we can
integrate it over an appropriate harmonic superspace measure
$d\m_{4,4}:=d^4 x\,du\,d^8\theta\,d^8\bar\theta$, where $du$ is
the usual Haar measure on the coset and where the odd variables
are $\th^{\a r'}:=\th^{\a i}(u^{-1})_i{}^{r'}$ and $ \bar
\th^{\adt}_{r}:=u_r{}^i \bar\th^{\adt}_i$. In order to obtain an
invariant the integrand must be $(4,4)$ G-analytic and must have
the right charge with respect to the $U(1)$ subgroup of the
isotropy group $S(U(4)\xz U(4))\sim U(1)\xz SU(4)\xz SU(4)$. The
only possible integrand with these properties which can be
constructed from $W$ is $W^4$; it gives the harmonic superspace
version of the three-loop counterterm in the form \cite{Hartwell:1994rp}

 \be
 I_{3-loop}=\int\,d\m_{4,4}\,W^4\, .
 \ee
Since $W\sim \th^4 C_{\a\b\c\d}+\bar\th^4 \bar
C_{\adt\bdt\cdt\ddt} +\ldots$, where $C$ is the Weyl spinor, it is
apparent that integrating over the odd variables will give the
square of the Bel-Robinson tensor.

All other possible invariants in linearised $N=8$ supergravity
which are invariant under $SU(8)$ and which involve integrating
over superspaces of smaller odd dimension than conventional
superspace were recently classified \cite{Drummond:2003ex}.
There are just two,
one in $(8,2,2)$ and one in $(8,1,1)$ harmonic superspace. They
both have the same schematic form as the three-loop invariant but
the measures and the definition of the superfield $W$ differ. They
correspond to five-loop and six-loop counterterms respectively:

 \bea
 I_{5-loop}&=&\int d\m_{2,2}\,W^4 \sim \int d^4x\, \del^4\,R^4 +\ldots\nn \\
 I_{6-loop}&=&\int d\m_{1,1}\,W^4 \sim \int d^4x\, \del^6\,R^4
 +\ldots \ .
 \eea

In order to discuss the relevance of these counterterms to
possible divergences in quantum supergravity one has to know how
much supersymmetry can be preserved in the quantum theory \cite{Grisaru:1982zh,Howe:1983}. This
means that we have to know the maximum number $M\leq 8$ of
supersymmetries that can be realised linearly in the off-shell
theory. If we stick to standard off-shell realisations, then only
$M=4$ is allowed for $N=8$ supergravity \cite{Rivelles:1982gn,Howe:1982mt}. This would then suggest
that counterterms should be integrals over sixteen odd
coordinates leading to a prediction of the first divergence
occurring at three loops. However, recent work by Bern {\sl et al}
has indicated that the coefficient of the three-loop counterterm
is not divergent \cite{Bern:1998ug}.

Some light can be shed on this apparent discrepancy by looking at
the maximally supersymmetric $N=4$ Yang-Mills theory which becomes
non-renormalisable in higher spacetime dimensions. In this case,
Bern {\sl et al} \cite{Bern:1998ug} also found improved UV behaviour compared to the
predictions of superspace power-counting which were again based on
the assumption that the best one can achieve off-shell is to
preserve half of the total number of supersymmetries \cite{Howe:1983jm}. However, in
the Yang-Mills case we know that there is an off-shell version
with $N=3$ supersymmetry available in harmonic superspace \cite{Galperin:1984bu}. If the
superspace power-counting is adjusted for this, then the new
predictions precisely match the calculational results \cite{Howe:2002ui}.

This line of reasoning suggests that there should be an off-shell
version of $N=8$ supergravity with $M=6$ supersymmetry
which would correspond to the first divergence appearing at five
loops. Results obtained in higher dimensions lend strong support
to this contention. In \cite{Bern:1998ug} it was shown that the maximal
supergravity theory in $D=7$ diverges at two loops where the
corresponding counterterm has a similar form to the $D=4$
five-loop invariant. The occurrence of this divergence indicates
that an off-shell version of the theory with six four-dimensional
supersymmetries should exist, whereas if one could preserve seven
such supersymmetries this divergence would not be allowed in
$D=7$. Unless there is a completely different mechanism at work in
four dimensions, it therefore seems that the $N=8$ theory is most
likely to diverge at five loops.

\section{Invariants in string theory and M-theory}

In discussing invariants corresponding to field-theoretic
counterterms it is enough to consider the linearised theory. In
string theory or M-theory, however, we are more interested in the
full non-linear expressions, and these are very difficult to find.
Information about particular terms in full invariants has been
derived from string scattering amplitudes and sigma model
calculations while some other hints have been obtained using
supersymmetry and arguments based on duality symmetries. An
important point to note is that the linearised invariants of the
type discussed in the preceding section cannot be generalised to
the non-linear case in any straightforward manner. This is because
the superspace measures do not exist in the interacting theory.
For example, $D=4, N=8$ supergravity does not admit an $(8,4,4)$
harmonic superspace interpretation in the full theory.

Invariants of the $R^4$ type are reasonably well understood in
$D=10, N=1$ supersymmetry where complete expressions are known for
the bosonic terms \cite{deRoo:1992zp}. There are two types of supergravity invariant.
The first is  a full superspace integral of the form \cite{Nilsson:1986rh}

 \be
 \int d^{10}x\,d^{16}\theta\,E f(\f) \sim \int
 d^{10}x\,e(\frac{d^4f}{d\f^4} R^4 + \ldots)
 \ee
where $E$ and $e$ denote the standard densities in superspace and
spacetime respectively, $\f$ is the dilaton superfield, and where
the $R^4$ terms appear in the combination
$(t_8t_8-{1\over8}\e_{10}\e_{10})R^4$ in the notation of, for
example, Peeters {\sl et al} \cite{Peeters:2000qj}, where these tensorial structures are explained. The
second type of invariant can be constructed starting from certain
Chern-Simons terms and so we shall refer to them as CS invariants.
In the next section we shall see how such invariants can be
explicitly constructed. There are two possible Chern-Simons terms
in supergravity, $B\wedge\tr R^4$ and $B\wedge(\tr R^2)^2$.
Details of all of these invariants up to quadratic order in
fermions have been computed \cite{deRoo:1992zp}.

The structures associated with the $N=1$ invariants are seen in
various combinations in the $N=2$ invariants. In IIA string theory
the pure curvature term in the tree level invariant has the form
$e^{-2\f}(t_8t_8-{1\over8}\e_{10}\e_{10})R^4$ , which resembles the
first type of $N=1$ invariant, while there is also a one-loop term,
required to cancel the five-brane anomaly, which arises from the CS
term $B\wedge X_8$, where $X_8= \tr R^4-\qu (\tr R^2)^2$. This is
believed to be associated with a pure $R^4$ term of the form
$(t_8t_8+{1\over8}\e_{10}\e_{10})R^4$. The $t_8^2$ terms have been computed from
string scattering \cite{Green:1981yb,Gross:1986iv,Sakai:1986bi}, the tree-level $\e^2$ term can be inferred from sigma 
model calculations \cite{Grisaru:1986px}, the one-loop $\e^2$ term is suggested 
by one-loop four-point amplitudes \cite{Kiritsis:1997em}, and the one-loop CS term is required for anomaly 
cancellation \cite{Vafa:1995fj}.

In IIB the linearised theory is described by a chiral superfield
$\F$ and there is a linearised invariant of the form $\int
d^{16}\theta\,\F^4 \sim (t_8t_8-{1\over8}\e_{10}\e_{10})R^4 +
\ldots$. This does not generalise to the full theory in any
straightforward manner \footnote{For an explicit attempt to do this see \cite{deHaro:2002vk}.}, however, due to the 
structure of the
superspace constraints of the full theory \cite{Howe:1983sr}. Nevertheless, with the use of
component supersymmetry, input from D-instanton results and
$SL(2,\bbZ)$ invariance some conjectures have been made about the
scalar structure of some terms in the invariant. Specifically,
each term appears multiplied by a function of the complex scalars
$(\t,\bar\t)$ which is a modular form under $SL(2,\bbZ)$ whose
weight is determined by the $U(1)$ charge of the term under
consideration \cite{Green:1998by}.

The $R^4$ invariant in M-theory has been discussed by various
groups using string theory \cite{Tseytlin:2000sf,deHaro:2002vk}, quantum
superparticles \cite{Green:1997as} and supergravity \cite{Russo:1997mk,Deser:2000xz}.
A straightforward supersymmetry
approach is rendered even more difficult by the fact that the
field strength superfield $W$ has dimension one so that one would
require an integral over eight odd coordinates of $W^4$ to obtain
$R^4$. One result that is known is that there must be a CS term to
cancel the anomaly of the M-theory five-brane \cite{Duff:1995wd}. This CS term has
the form $C_3\wedge X_8$, where $C_3$ is the three-form gauge
field in the theory.

An alternative approach is to investigate the modified equations
of motion rather than trying to compute the invariant directly.
Corrections to the equations of motion can be understood as
deformations of the on-shell superspace constraints, and the
consistency conditions that the Bianchi identities place on these
constraints lead, when one takes the into account the possibility
of field redefinitions of the underlying superspace potentials, to
a reformulation of the problem in terms of certain superspace
cohomology groups. This cohomology, called spinorial cohomology,
has been studied for various theories in the literature including
M-theory considered as a deformation of $D=11$ supergravity \cite{Cederwall:2001dx,Howe:2003cy}. In
certain circumstances spinorial cohomology coincides with pure
spinor cohomology which has been used to give a new
spacetime supersymmetric formulation of superstring theories \cite{Berkovits:2000nn}.

In the final section we shall investigate M-theoretic $R^4$ terms
in a superspace setting by looking at the Bianchi identities in
the presence of the five-brane anomaly cancelling term. It will be
argued that the CS term gives rise to a unique invariant which is
both supersymmetric and consistent with the anomaly and it will be
shown how this invariant can be constructed. The deformations of
the superspace constraints are driven by the anomaly term and can
be found systematically, at least in principle.

\section{M-theory in superspace}

The component fields of $D=11$ supergravity are the elfbein
$e_m{}^a$, the gravitino, $\psi_m{}^\a$, and the three-form
potential $c_{mnp}$ \cite{Cremmer:1978km}. In the superspace formulation of the theory \cite{Cremmer:1980ru}
the first two appear as components of the supervielbein $E_M{}^A$,
whereas the third is a component of a superspace three-form
potential $C_{MNP}$.\footnote{Notation: latin (greek) indices are
even (odd), capital indices run over both types; coordinate
(tangent space) indices are taken from the middle (beginning) of
the alphabet.} We have

 \bea
 E^a=dz^M E_M{}^a &=& \begin{cases} dx^m E_m{}^a  &=dx^m(e_m{}^a + O(\th))
 \\ d\th^{\m}\, E_{\m}{}^a  &=d\th^{\m}(0 + O(\th) ) \end{cases}
 \nn\\
 &&\nn\\
 E^{\a}=dz^M E_M{}^{\a} &=& \begin{cases} dx^m E_m{}^{\a} &=dx^m (\psi_m{}^\a + O(\th))\\
 d\th^{\m}\, E_{\m}{}^{\a} &=d\th^{\m}(\d_\m{}^\a + O(\th))\end{cases}
 \eea
 while $C_{mnp}(x,\th=0)=c_{mnp}$.

 The structure group is taken to be the Lorentz group, acting
 through
 the vector and spinor representations in the even and odd sectors
 respectively.
 We also introduce a connection one-form $\O_A{}^B$ which takes its
 values in the Lie algebra of the Lorentz group and define the
 torsion and curvature in the usual way:

  \bea
 T^A &=&DE^A:=dE^A + E^B \O_B{}^A=\half E^C E^B T_{BC}{}^A \nn\\
 R_A{}^B &=&d\O_A{}^B +\O_A{}^C\O_C{}^B=\half E^D E^C R_{CD,A}{}^B
 \eea
 From the definitions we have the Bianchi identities $DT^A=E^B
 R_B{}^A$ and $DT^A=0$. The assumption that the structure group is
 the Lorentz group implies that $R_a{}^{\b}=R_{\a}{}^b=0$ while

 \be R_{\a}{}^{\b}={1\over4} (\c^{ab})_{\a}{}^{\b} R_{ab} \ee

 The equations of motion of supergravity in superspace are implied
 by constraints on the torsion tensor. In fact, it is enough to
 set

  \be
  T_{\a\b}{}^c=-i(\c^c)_{\a\b}
  \ee
 to obtain this result \cite{Howe:1997rf}. The only other components of the torsion
 which are non-zero are

  \be T_{a\b}{}^{\c}= -{1\over36}\left((\c^{bcd})_{\b}{}^{\c}
W_{abcd} +{1\over8} (\c_{abcde})_{\b}{}^{\c} W^{abcd}\right) \ee
 and $T_{ab}{}^{\c}$ whose leading component can be identified as
 the gravitino field strength. Given these results one can
 construct a superspace four-form $G_4$ which is closed and whose
 only non-zero components are

  \be
  G_{\a\b cd}=-i(\c_{cd})_{\a\b},\qquad G_{abcd}=W_{abcd}
  \ee
The only independent spacetime fields described by these
 constraints are the physical fields of supergravity; their
 field strengths are the independent components of the superfield
 $W$.

 Instead of deducing the existence of a four-form $G_4$ we can
 include it from the beginning. In this case one can show that
 there is a stronger result, namely that imposing the constraint

  \be
  G_{\a\b\c\d}=0
  \ee
 is sufficient to imply the supergravity equations of motion \cite{Cederwall:2001dx,Berkovits:2002uc,Howe:2003cy}.
 Either the geometrical approach or the four-form approach can be
 used as a starting point for investigating deformations using
 spinorial cohomology, but we shall choose yet
 another route by introducing a seven-form field strength $G_7$ as
 well.\cite{Howe:2003cy} We then have the coupled Bianchi identities

  \be
  d G_4=0\, , \qquad d G_7=\half (G_4)^2 + \b X_8
  \ee
 where we have included the anomaly term and where $\b$ is a
 parameter of dimension $\ell^6$.

 We shall work to first order in $\b$ which means that we can use
 the supergravity equations of motion in computing $X_8$, in other
 words, $X_8$ is a known quantity. The Bianchi identities can be
 solved systematically and all of the components of $G_4, G_7, T$
 and $R$ can be found. Note that the only component of any of these
 tensors which can be zero at order $\b$ is $G_{\a_1\ldots \a_7}$.
 However, this solution might not be unique as there could be
 solutions of the homogeneous equations (i.e. without the $X_8$
 term). In principle this question could be tackled using spinorial
 cohomology but we shall study it indirectly by looking at the
 action.

 We briefly describe how one can construct a superinvariant from
 any CS invariant. If we have a theory in $D$ spacetime dimensions
 formulated in superspace an invariant can be constructed if we
 are given a closed superspace $D$-form $L_D$ \cite{D'Auria:1982pm,Gates:1997ag}. This invariant is

  \be
  I=\int d^D x\, \e^{m_1\ldots m_D} L_{m_1\ldots m_D}(x,\th=0)
  \ee
 Under a superspace diffeomorphism generated by a vector field $v$

  \be
  \d L_D={\mathcal L}_v L_D=d(\i_v L_D) + \i_v dL_D=d(\i_v L_D)\ .
  \ee
 Identifying the $\th=0$ components of $v$ with the spacetime
 diffeomorphism and local supersymmetry parameters we see from this equation
 that the above integral is indeed invariant.

 In some situations, notably when we have CS terms available, we
 can easily construct such closed $D$-forms \cite{Bandos:1995dw,Howe:1998ts}. Suppose there is a
 closed $(D+1)$-form $W_{D+1}=d Z_D$ where $Z_D$ is a potential
 $D$-form which we are given explicitly. We can always write
 $W_{D+1}=d K_D$, where $K_D$ is a globally defined $D$-form,
 because the cohomology of a real supermanifold is equal to that of
 its body and this is trivial in degree $D+1$ in $D$
 dimensions. If we set $L_D=K_D-Z_D$ then  $L_D$ is
 closed and hence gives a rise to a superinvariant using the
 construction described above. Any CS term gives rise to a superinvariant in this
 manner.

 In M-theory the appropriate forms are

 \bea
 W_{12}&=&\half G_4^3 + 3\b G_4 X^8\nn \\
 Z_{11}&=&C_3(\half G_4^2 + 3\b X_8)
 \eea
 The invariant constructed from these forms will include the
 anomaly-cancelling CS term and superpartners which will be of
 $R^4$ type. The purely bosonic terms, aside from the CS term
 itself, come from $K_{a_1\ldots a_{11}}$, and these are the most
 difficult to compute. The easiest term to calculate is the lowest
 non-vanishing term which is $K_{abc\d_1\ldots \d_8}$. It gives
 rise to terms with eight gravitinos in the action.

 To this invariant we could add any invariant coming from a closed
 form $L_{11}$. It has been argued on (purely algebraic) cohomological grounds
 (subject to some assumptions) that
 there are no such terms \cite{Howe:2003cy}. If this is the case we would
 conclude that the $R^4$ invariants in eleven
 dimensions  come from CS terms. In principle there could be two of
 these, since we have both $\tr R^4$ and $(\tr R^2)^2$, but
 M-theoretic considerations imply that we have to choose the
 linear combination which appears in $X_8$.

 Finding the explicit form of this invariant is an extremely
 difficult task. One could start either by solving the $G_7$
 Bianchi identities, or one could try to find $K_{11}$ directly by
 solving the equation $W_{12}=d K_{11}$. This is currently under investigation.

\section*{Acknowledgements}

The recent results on counterterms reported here were obtained with Kelly Stelle whom I thank for many discussions on 
this subject. I thank Ulf Gran and Dmitris Tsimpis for ongoing discussions on solving the modified Bianchi identities.

 \end{document}